# BL Lac: A New Ultrahigh-Energy Gamma-Ray Source

Yu. I. Neshpor, N. N. Chalenko, A. A. Stepanian, O. R. Kalekin, N. A. Jogolev,
V. P. Fomin, and V. G. Shitov

*Crimean Astrophysical Observatory, p/o Nauchnyĭ, Crimea*

Received May 3, 2000

**Abstract**—The active galactic nucleus BL Lac was observed with the GT-48 atmospheric Cherenkov detector of the Crimean Astrophysical Observatory from July 23–September 1, 1998, in order to search for ultrahigh-energy gamma-ray (>1 TeV) emission. The object was in the field of view of the detector for more than 24 hours. The source was detected with a high level of confidence (7.2 σ), with a flux equal to $(2.1 \pm 0.4) \times 10^{-11}$ photons cm$^{-2}$ s$^{-1}$. © *2001 MAIK "Nauka/Interperiodica".*

## 1. INTRODUCTION

Interest in active galactic nuclei (AGN) among cosmic-ray specialists rose sharply after the Compton Gamma-Ray Observatory (CGRO) demonstrated that a large fraction of sources of high-energy gamma-rays (>100 MeV) were identified with AGN [1]. These galaxies were subsequently observed with ground-based instruments designed to detect gamma-rays with energies of $10^{11}$–$10^{12}$ eV. Two AGN were found to be sources of ultrahigh-energy gamma-rays using the 10-m ground Cherenkov detector of the Whipple Observatory in the United States: the Makarian galaxies Mrk 421 and Mrk 501 [2]. At the Crimean Astrophysical Observatory (CrAO), high-energy gamma-rays were also detected from the blazar 3C 66A [3], and classified, like Mrk 421 and Mrk 501, as a BL Lac object; BL Lac is the prototype of this relatively small group of AGN.

One of the main properties of BL Lac objects is variability in their brightness, which can reach $4^m$–$5^m$ in the optical (corresponding to luminosity variations by factors of 100). These objects all have appreciable radio emission, which is also, as a rule, variable. BL Lac objects are also characterized by featureless or nearly featureless optical continua, without strong emission lines. The power-law nature of their spectra and the strong polarization of their radiation, which reaches 30–40%, testify that we are observing synchrotron radiation. The characteristic variability timescales—weeks to months—indicate that the dimensions of the radiating volumes in BL Lac objects are typically of the order of $10^{16}$ cm. The nature of the processes taking place in these galactic nuclei suggest that acceleration of highly energetic particles should occur there, which could be accompanied by the radiation of ultrahigh-energy (UHE; $E > 10^{11}$ eV) gamma-rays. The redshift of BL Lac itself is 0.07, corresponding to a distance of 280 Mpc.

The galaxy BL Lac has been observed using the ground-based UHE gamma-ray detectors of the Whipple Observatory in the 1970s [4] and at the CrAO [5]. Stepanian *et al.* [5] obtained an upper limit for the flux $F < 1.1 \times 10^{-10}$ photons cm$^{-2}$ s$^{-1}$ for an energy threshold $2.2 \times 10^{12}$ eV. The 1971 observations of the American group [4] yielded an upper limit of $F < 1.2 \times 10^{-10}$ photons cm$^{-2}$ s$^{-1}$ for energies $E > 2.5 \times 10^{11}$ eV. BL Lac was again observed by the Whipple Observatory in October–November 1994, for a total time of 162 min. These observations yielded an upper limit on the flux of $1.4 \times 10^{-11}$ photons cm$^{-2}$ s$^{-1}$ [6].

During this time, BL Lac was observed by the CGRO. Observations before January 1995 indicated a flux of $1.4 \times 10^{-7}$ photons cm$^{-2}$ s$^{-1}$ [7], corresponding to a 2.4 σ detection. However, in January–February 1995, the flux rose to $(4.0 \pm 1.2) \times 10^{-7}$ photons cm$^{-2}$ s$^{-1}$, providing evidence for variability in the gamma-ray emission from BL Lac. Studies of BL Lac at the CrAO were renewed in 1998 using the second-generation GT-48 gamma-ray telescope [3, 8].

## 2. BRIEF DESCRIPTION OF THE GT-48 GAMMA-RAY TELESCOPE

Gamma rays with ultrahigh energies (~$10^{11}$ eV) cannot reach the surface of the Earth. They interact with the nuclei of atoms in the atmosphere, forming so-called broad air showers (BASs) consisting of high-energy electrons and positrons. The charged particles of BASs emit Cherenkov radiation at optical wavelengths at small angles (0.5°–1°) to the direction of motion of the original gamma ray, making it possible to not only detect the presence of the gamma ray but also determine the direction toward its source. However, charged cosmic-ray particles also give rise to Cherenkov flares in the Earth's atmosphere. These are very similar to flares due to gamma-ray sources, leading to the main difficulty in detecting and studying these sources. Nevertheless, differences between these two types of flares do exist, and multielement light-collecting chambers that reconstruct images of Cherenkov flares can be used to eliminate the overwhelming majority of events resulting from charged cosmic-ray particles.

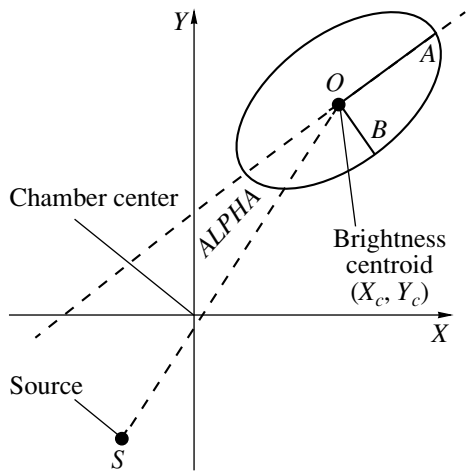

**Fig. 1.** Graphical depiction of flare parameters. The segment *OS* corresponds to the parameter *DIST*.

The GT-48 telescope equipped with multichannel imaging chambers began to operate at the CrAO in 1989. The instrument consists of two altitude–azimuth mounts (sections)—northern and southern—separated by 20 m and located at a height of 600 m above sea level. Six parallel telescopes are mounted on each section. The optics for each telescope include four 1.2-m mirrors with a common focus. The mirrors of the four telescopes have a focal length of 5 m. Chambers with 37 photomultipliers are placed at each focus. These are used to reconstruct the images of Cherenkov flares at visible wavelengths (300–600 nm). There is a conical light guide at the entrance to each photomultiplier. The mean diameter of the light guide entry window is 0.4°. The field of view of the entire light-collecting apparatus is 2.6°. The signals from the cells of the four light receivers are linearly summed, and arrive at a amplitude-code converter via the 37 channels. In this way, a discrete image of the Cherenkov flare is obtained, which can be stored on a computer. Flares are recorded only if the amplitudes of signals coincident in time exceed a specified threshold in any two of the 37 cells. The time resolution for evaluating coincidence is 15 ns.

Two other telescopes with a focal length of 3.2 m are intended for observations at ultraviolet wavelengths, from 200 to 300 nm. Although the fluxes from BASs are weak in this part of the spectrum, detecting this radiation can aid in distinguishing showers due to cosmic gamma-ray sources against the background of showers due to charged cosmic-ray particles. In this sense, the GT-48 telescope is unique. We will discuss the use of ultraviolet emission to distinguish showers due to gamma-ray sources in more detail below. The total area of both mirrors is 54 m$^2$. The effective detection threshold energy of gamma rays is 1.0 TeV. More detailed descriptions of the GT-48 telescope can be found, for example, in [3, 8].

## 3. OBSERVATIONS AND DATA ANALYSIS

Observations of the object BL Lac ($\alpha = 22^h02^m43^s$, $\delta = 42°16'40''$) were carried out in July–August 1998 using the two parallel sections in a coincidence regime with a time resolution of 100 ns; i.e., we recorded only events that were observed simultaneously on both sections to within this time resolution. Each section tracked the source and also made observations of the background, with the time between these observations equal to 40 min. The background recordings were performed at the same azimuths and zenith angles as those for the source, and preceded the source observations. In all, there were 63 sessions, with the duration of the source observations in each being 35 min. The zenith angle did not exceed 30°. We did not include sessions carried out under poor weather conditions in the reduction. As criteria for the selection of sessions, we used the dispersion of the count rate per minute during the session and the mean count rate. We excluded sessions in which the dispersion of the count rate differed from the theoretical value by more than two $\sigma$ or the count rate was lower than half the maximum value for a given zenith angle. After this selection, 42 sessions remained, corresponding to 24 hours and 30 minutes of observations of BL Lac. Further, we excluded events for which a maximum of the amplitude-code transformation was reached in at least one of the channels, and also events for which the maximum amplitude was detected in one of the outer channels of the light receiver. After this preliminary selection, 30340 source events and 29489 background events remained for further analysis. The difference in the numbers of flares recorded for source and background observations is $N_\gamma = 851 \pm 245$, where 245 is the statistical error $\sigma = \sqrt{N_s + N_b}$, and $N_\gamma$ is the number of Cherenkov flares from gamma rays.

The resulting data were subject to further reduction, to analyze the digitized flare images using formal mathematical methods. We computed the first and second moments of the brightness distributions, from which we derived the parameters of the Cherenkov flares: the coordinates of the center of the brightness distribution $X_c$ and $Y_c$, the effective length $A$, the effective width $B$, and angle $\varphi$ characterizing the direction of maximum extension of the flare image, i.e., its orientation (Fig. 1). These moments were computed for cells with signals greater than some threshold value [9]. The parameters of a flare (event) recorded simultaneously on each section were determined independently using the data acquired for each section, so that each event has two values for each parameter, which we denote "1" and "2" for the northern and southern sections, respectively.

It is known that electrons from proton and nuclear showers ($p$ showers, BASs) of a given energy, on average, penetrate to appreciably larger depths in the Earth's atmosphere than do electrons from gamma-ray showers of the same energy. This means that Cherenkov flares from $p$ showers contain a relatively larger fraction of ultraviolet radiation compared to flares formed by gamma rays. As a parameter characterizing the relative content of ultraviolet radiation, we use the logarithm of the ratio of the flare amplitude in the ultraviolet and the total flare amplitude in the visible: UV.

**Table 1.** Number of recorded and selected events

| Selection method | Number of events in the source direction | Number of events in the background direction | Difference | Signal-to-noise, standard deviations |
|---|---|---|---|---|
| Without selection | 30340 | 29489 | 851 | 3.5 |
| Selection in coordinate-independent parameters | 1166 | 881 | 285 | 6.3 |
| Selection in coordinate-dependent parameters | 259 | 119 | 140 | 7.2 |

Flares from charged particles have larger angular sizes. In addition, they are distributed isotropically, while images of flares formed by gamma rays are extended in the direction toward the source. The angle between the direction toward the source from the flare center and the major axis of the image ellipse—the parameter *ALPHA*—can also be used to distinguish gamma-ray events. Using these differences between images of Cherenkov flares associated with cosmic-ray particles and gamma rays, it is possible to eliminate 99% of background events while leaving an appreciable fraction of gamma-ray events.

Events whose parameter values did not fall in a specified range were excluded from consideration. We selected those parameter values for which the effect in terms of standard deviations—that is, $N_s - N_b/\sqrt{N_s + N_b}$ (the signal-to-noise ratio)—was maximum. Here, $N_s$ and $N_b$ are the number of selected events for the source and the background, respectively. In the selection, first and foremost, we considered the total amplitude of flare *V*. Flares with small amplitudes were excluded from further consideration, since their parameters had large errors. It is also known that images of flares associated with proton showers have complex shapes, and can have several maxima or be fragmented, while flares from gamma-ray showers are compact. The parameter used to characterize the shapes of the flares is denoted IPR, and it is assigned the value 0 for compact images. We used the effective length *A* and width *B* of the flare images as parameters to distinguish gamma-ray showers against the background of charged-particle showers.

In this way, we excluded events from further consideration if at least one of the following conditions was satisfied: $V(1) < 100$, $V(2) < 100$, $IPR(1) \neq 0$, $IPR(2) \neq 0$, $A(1) > 0°.30$, $A(2) > 0°.30$; $B(1) > 0°.175$, $B(2) > 0°.175$, $UV(1) > 1.1$, $UV(2) > 0.8$. The parameters *V*, IPR, *A*, *B*, and UV do not depend on the position of the source relative to the flare; i.e., they are coordinate-independent. The application of these parameters made it possible to increase the signal-to-noise ratio to 6.3 σ. Our results are presented in Table 1.

It is possible to refine or even accurately determine the direction of a gamma-ray source using coordinate-dependent parameters, such as *ALPHA*—the orientation of the flare image relative to the direction of a proposed (trial) gamma-ray source—and *DIST*, which is numerically equal to the angular distance between the center of the flare brightness distribution and the direction toward the trial source (see Fig. 1). In this case, among Cherenkov flares selected based on coordinate-independent parameters, we identified gamma-ray-like events for which $ALPHA(1) < 30°$, $ALPHA(2) < 30°$, $0°.25 < DIST(1) < 0°.95$, and $0°.25 < DIST(2) < 0°.95$. The last column of Table 1 presents the results of this selection (selection based on coordinate-dependent parameters).

## 4. RESULTS

Thus, as a result of the selection procedure described above, we have been able to separate out reliable detections of gamma-ray fluxes. The count rate for the recorded gamma rays was $0.095 \pm 0.013$ photons/min. In order to determine the corresponding flux, we must take into account the effective area of the detector and determine what fraction of the gamma rays remained after the selection procedure. This can be done via theoretical Monte-Carlo studies of the gamma-ray showers. This method was used to numerically model the development of broad atmospheric showers and their detection by the GT-48 telescope [10].

This task was carried out in two stages. We first obtained all of the parameters needed to model the detection of gamma rays by the GT-48 telescope by comparing the results of our simulations of the recording of the cosmic-ray background with the observational data. Further, we simulated the recording of gamma rays and computed the corresponding fluxes. The simulations were carried out using code written by A.V. Plyasheshnikov (see [11]). Given a type of initial particle and the angle for its entry into the atmosphere, the code computes the number of Cherenkov photons from a BAS at the height of the detector.

To model recording of the background, we used the results of computations of showers due to protons and helium nuclei. These form the vast majority of cosmic rays at ultrahigh energies. The differential energy spectral indices for protons and helium are 2.75 and 2.62, respectively [12]. In all, we modeled 133 610 proton and

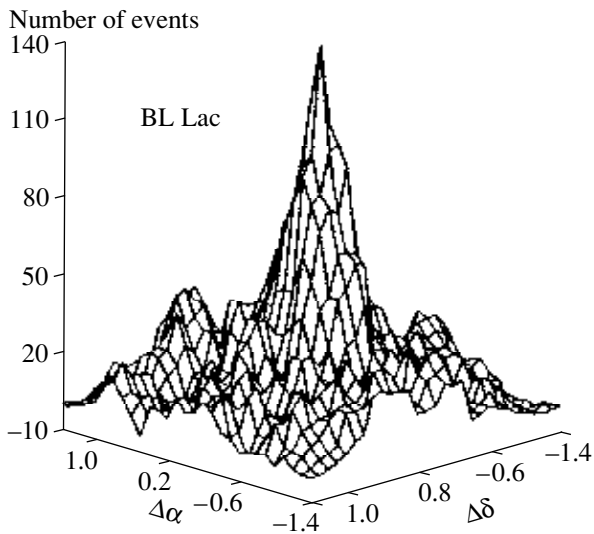

**Fig. 2.** Stereoimage of a "map" of the distribution of arrival directions of gamma rays. The center of the "map" coincides with the coordinates of BL Lac; $\Delta\delta$ is the deviation from the source in declination and $\Delta\alpha$ the deviation in right ascension (in degrees).

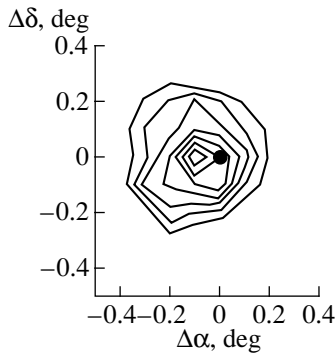

**Fig. 3.** Isophotes of the distribution of gamma-ray arrival directions. The notation is the same as in Fig. 2. The isophote step is ten events. The first value is 70 events. The black circle at the center shows the optical position of BL Lac.

56 670 helium events at energies from 0.4–70 TeV. The energy distribution of the events was chosen to minimize the statistical error in all energy ranges. Therefore, for each event, we computed a weight taking into account the differential flux and spectral index [12]. Further, we modeled the recording of background events by the gamma-ray telescope. The radio-technological threshold in the model was chosen to provide a count rate for background events equal to 72 min$^{-1}$; i.e., the value obtained in the observations.

The last required model parameter is the coefficient for the conversion of photo-electrons into a discrete digital shape. We derived this coefficient by requiring that the number of events for which at least one of the channels was saturated (a maximum in the analog-code transformation) be ~10% of the number of recorded events. This ratio was likewise obtained from the observational data. Thus, using model parameters derived by comparing modeling results with observations of background cosmic rays, we were able to model the process of recording gamma-ray events. In all, we modeled 5519 gamma events at energies 0.4–20 TeV with a differential energy spectral index of 2.4. The effective energy threshold for recording the gamma rays was 1.0 TeV. All events were parametrized in the same way as the observations. We selected model gamma events using the same criteria as in our analysis of the data for BL Lac. After this selection, 21.6% of the gamma events remained. This indicates that, since the count rate for the selected observed gamma events was 0.095 min$^{-1}$, the initial rate should be 0.44 min$^{-1}$. This count rate corresponds to a flux of gamma rays with energies >1.0 TeV equal to $(2.1 \pm 0.4) \times 10^{-11}$ photons cm$^{-2}$ s$^{-1}$.

We applied a trial-source method to determine the direction of the gamma-ray flux [13–15]. This method is based on the fact that the images of Cherenkov flares from gamma rays are oriented toward the source in the focal plane of the telescope, while the major axes of the image ellipses for $p$ showers are, to first approximation, oriented uniformly in all directions. Therefore, if we select flares, adopting as the direction toward the source an arbitrary point in the focal plane with coordinates $(X_i Y_j)$, and then make a selection according to coordinate-dependent parameters, the number of selected $p$ showers will not depend on the position of the trial source, while the number of selected gamma events will have a maximum when the trial source coincides with the actual source. If we now subtract from the number of flares obtained toward a trial source the number of background flares obtained for those same coordinates, we obtain the distribution of gamma rays over the field of view of the light receiver as a function of the position of the trial source $N(X_i Y_j)$; i.e., a "map" (two-dimensional histogram) of gamma-like events $N_\gamma$. In this way, we can find the position of the true source of gamma rays. Figure 2 presents a two-dimensional histogram obtained using the coordinate-dependent criteria *ALPHA* and *DIST*.

Figure 3 shows isophotes of $N_\gamma$. The first isophote corresponds to 50% of the maximum value of $N_\gamma$, which is equal to $(140 \pm 19.4)$ events $(7.2\,\sigma)$ and corresponds to the point with coordinates $X = -0.1$, $Y = 0.0$. During the observations, the center of the chambers were pointed at BL Lac. It is clear from Fig. 3 that the coordinates of the UHE gamma-ray source coincide well with the optical coordinates for BL Lac.

## 5. TEMPORAL VARIATIONS

Optical observations of BL Lac have been conducted for many years. BL Lac was first discovered as a stellar object in 1929. Observations over long timescales demonstrated that variations in its brightness reach $5^m$ (corresponding to luminosity variations of a factor of 100) [16]. Quasi-periodic fluctuations with periods of 0.8 and 0.6 yrs have been detected, as well as more rapid brightness variations.

**Table 2.** Powers of several BL Lac objects

| Object | Distance, Mpc | log L, [erg/s]/reference | | |
|---|---|---|---|---|
| | | optical | high energy | ultrahigh energy |
| BL Lac (1998) | 280 | 44.8 [17] | (1995) 44.9 [20] | 44.5 [this paper] |
| Mrk 501 | 160 | 44.4 [19] | – | 44.5 [22] |
| 3C 66A | 1800 | 46.4 [19] | 46.2 [20] | 46.3 [23] |
| Crab | 0.002 | 36.5 [24] | 35.3 [20] | 34.0 [21] |

During the period for our observations (July–August 1998), the visual brightness of BL Lac varied from $14.6^m$ to $13.5^m$ [17]. Rapid variations with amplitude $0.5^m$ were noted. Unfortunately, the detailed temporal variations in the visual luminosity of BL Lac presented in [17] do not coincide in time with our observations.

Figure 4 presents the mean gamma-ray count rate per minute over an observing night. We conclude from this figure that the UHE gamma-ray flux is time variable. During the optical observations (July–August 1998), the total optical energy flux varied from $3.8 \times 10^{44}$ to $1.0 \times 10^{45}$ erg/s. During this time, the mean total energy flux in UHE gamma rays was $3.2 \times 10^{44}$ erg/s.

It is interesting to compare the power emitted by BL Lac objects in various energy ranges. We composed Table 2 from data available in the literature, assuming that the radiation at all energies is isotropic. For comparison, we also present in Table 2 the power of the Crab Nebula. We can see that, for the gamma-ray source BL Lac, as well as the objects Mrk 501 and 3C 66A, the total powers in the optical, at high energies, and at ultrahigh energies are quite comparable. Unfortunately, EGRET did not conduct any observations of BL Lac during the period under consideration, and we therefore present measurements for 1995 in the table. Note that the flux of gamma rays with energies >100 MeV greatly increased during the optical flare of July 19, 1997 [18]. At maximum light, the brightness of BL Lac reached almost $12^m$ (a luminosity of $5 \cdot 10^{45}$ erg s$^{-1}$), while the gamma-ray flux at energies >100 MeV reached $5 \cdot 10^{-6}$ phot. cm$^{-2}$ s$^{-1}$ ($5 \cdot 10^{46}$ erg s$^{-1}$). The energy spectrum of the photons in July 1997 was appreciably harder than in more quiescent periods: the spectral index was $1.7 \pm 0.1$, as opposed to $2.3 \pm 0.3$. If we suppose that the spectral slope during the July 1997 flare was preserved to $10^{12}$ eV, the power in UHE gamma rays should have been no lower than $5 \times 10^{46}$ erg/s, and should have exceeded the power in the optical (Table 2) by a order of magnitude.

## 6. CONCLUSION

Thus, we conclude that BL Lac is a source of UHE gamma rays. It is likely that the flux of ultrahigh-energy gamma rays is variable. We expect that this flux rises sharply in periods of enhanced optical brightness. The spectra of BL Lac, Mrk 501, and 3C 66A are similar. It would be very desirable to observe BL Lac simultaneously in a number of different wavebands across the electromagnetic spectrum: from radio to UHE gamma energies.

## ACKNOWLEDGMENTS

The authors thank S.G. Kochetkov and Z.N. Skiruta for help in reducing the data and preparing the article.

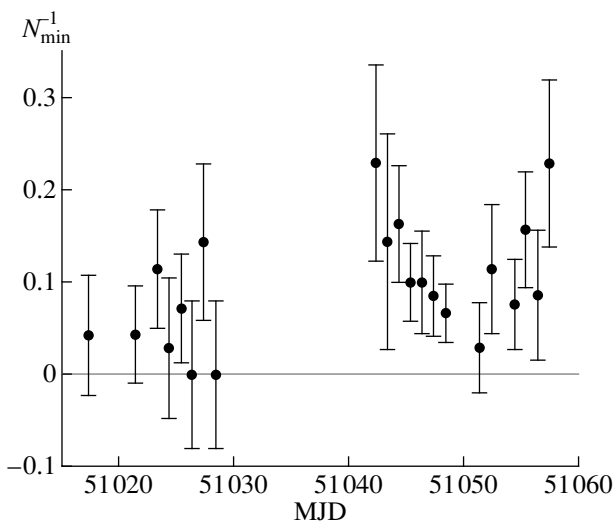

**Fig. 4.** Time behavior of the gamma-ray flux (mean for an observing night). The errors shown are statistical.

*Translated by D. Gabuzda*